\documentclass[twocolumn,preprintnumbers,amsmath,amssymb,pra,floatfix]{revtex4-1}

\usepackage{graphicx}
\usepackage{dcolumn}
\usepackage{bm}

\begin{document}

\title{Magnetic Anistropy due to the Casimir Effect}

\author{G. Metalidis}
\affiliation{Institut f\"{u}r Theoretische Festk\"{o}rperphysik and DFG Center for Functional Nanostructures (CFN), Karlsruhe Institute of Technology (KIT), D-76131 Karlsruhe, Germany}
\author{P. Bruno}
\affiliation{European Synchrotron Radiation Facility, BP 220, F-38043 Grenoble Cedex, France}

\date{\today}

\begin{abstract}
We consider the Casimir interaction between a ferromagnetic and a non-magnetic mirror, and show how the Casimir effect gives rise to a magnetic anisotropy in the ferromagnetic layer. The anisotropy is out-of-plane if the non-magnetic plate is optically isotropic.  If the non-magnetic plate shows a uniaxial optical anisotropy (with optical axis in the plate plane), we find an in-plane magnetic anisotropy. In both cases, the energetically most favorable magnetization orientation is given by the competition between polar, longitudinal and transverse contributions to the magneto-optical Kerr effect, and will therefore depend on the interplate distance. Numerical results will be presented for a magnetic plate made out of iron, and non-magnetic plates of gold (optically isotropic), quartz, calcite and barium titanate (all uniaxially birefringent).
\end{abstract}

\maketitle

\section{Introduction}
Even 60 years after its discovery, the Casimir effect continues to fascinate~\cite{Casimir1948}. Originally calculated as an attractive force between two neutral metallic plates due to the zero-point energy of the electromagnetic field, the Casimir effect nowadays plays a role in fields as diverse as gravitation, cosmology, atomic and molecular physics, and quantum field theory~\cite{Bordag2001}. From a condensed matter perspective, research has been mainly preoccupied with the inclusion of several contributions that are present in a real experimental setup, e.g., boundary roughness, shape and geometry of the mirrors, non-zero temperature, and dielectric properties of the boundaries (a thorough review can be found in Ref.~\cite{Bordag2001}). With the theory in place, several recent experiments have verified the existence of the Casimir force~\cite{Lamoreaux1997, Mohideen1998, Chan2001, Bressi2002, Decca2003, Chan2008, Munday2009, Klimchitskaya2006}).

Whereas most publications deal with optically isotropic mirrors, the Casimir effect between anisotropic mirrors is a rather recent topic. The van der Waals interaction between uniaxial anisotropic media of infinite thickness has been considered in Ref.~\cite{Barash1973} and formed the starting point for predictions of a torque between two uniaxial birefringent plates~\cite{vanEnk1995, Munday2005, Shao2005}.  The zero-point energy depends on the relative orientation of the plates which gives rise to a torque that seeks to align their principal axes. Similar effects are present for plates in which the optical anisotropy is induced by corrugations~\cite{Rodrigues2006}. Furthermore, in a search for repulsive instead of attractive Casimir forces, and stimulated by the advent of so-called magneto-dielectric metamaterials (which have a non-negligible magnetic response on top of the electric one), the effects of anisotropy of the magnetic permeability on the Casimir effect were studied in Refs.~\cite{Rosa2008, Deng2008}. The Casimir force between  ferromagnetic mirrors, which of course cannot be described by a magnetic permeability, has been treated by us in a recent publication~\cite{Bruno2002, Metalidis2002}. In this case, the boundary conditions for the zero-point electromagnetic field are influenced by the magneto-optical Kerr effect, and we have found that this gives rise to a magnetic contribution to the Casimir force.

In the present paper, the force between a single ferromagnetic mirror opposite of a non-magnetic one is considered.  Interestingly, we find that the Casimir effect will give rise to a \textit{magnetic} anisotropy in the ferromagnet. We will first study a setup where the non-magnetic mirror is optically isotropic, and observe an out-of-plane magnetic anisotropy in the ferromagnetic layer. For this setup, the anisotropy energy arising from the Casimir effect is orders of magnitude smaller than the surface anisotropy and magnetocrystalline anisotropy. Therefore we will propose another scheme where the non-magnetic mirror is made of a uniaxially birefringent material (with optical axis in the plane). The Casimir effect then induces an in-plane magnetic anisotropy that can be detected more easily. In both cases, the magnetic orientation with the lowest energy results from a competition between several contributions to the magneto-optical Kerr effect, and will depend on the distance between the plates.

\section{General Theory}
The Casimir energy (per unit area) between two plates $A$ and $B$ can be expressed in terms of the reflection coefficients of the plates as~\cite{Jaekel1991, Metalidis2002}
\begin{widetext}
\begin{equation}
{\cal E}=\frac{\hbar}{(2\pi)^{3}} \int_{0}^{+\infty} {\rm d}
k_{\perp}\,k_{\perp}  \int_{0}^{2\pi} {\rm d}\varphi
\int_{0}^{k_{\perp}c} {\rm d}\omega\, \mbox{Re Tr} \ln[1-{\sf
R}_{A}({\rm i}\omega,{\rm i}k_{\perp},\varphi) {\sf R}_{B}({\rm
i}\omega,{\rm i}k_{\perp}, \varphi){\rm e}^{-2k_{\perp}D}],
\label{eq:CasEnergy}
\end{equation}
\end{widetext}
where $\hbar \omega$ is the electromagnetic field frequency,  $k_{\perp}$ is the component of the wavevector perpendicular to the mirrors, and $\varphi$ is the angle between the incidence plane and the $X$-axis. Furthermore, ${\sf
R}_{A/B}$ are $2\times2$ matrices containing the reflection coefficients of the plates (evaluated at imaginary wavevector and frequency):
\begin{equation}
{\sf R}_{A(B)}=
\left(\begin{array}{cc}
 r_{ss}^{A(B)} & r_{sp}^{A(B)} \\
 r_{ps}^{A(B)} & r_{pp}^{A(B)}
 \end{array}\right).
\end{equation}
The index $s$ (resp. $p$) corresponds to a polarization with the electric field perpendicular (resp. parallel) to the incidence plane. We adopt the convention that the $p$ axis remains unchanged upon reflection.

In this paper, we will consider one of the plates to be ferromagnetic. The reflection coefficients are then influenced by the magneto-optical Kerr effect, and depend on the magneto-optical constant $Q$. Since $Q$ is small, we will only keep terms up to first order in $Q$, in which case the reflection matrix can be written as $R = R_0 + \Delta R$, with
\begin{equation} \label{eq:reflmatrixR0}
R^0 = \left[ \begin{array}{cc}
r_{ss} &  0 \\
0 & r_{pp}
\end{array} \right]
\end{equation}
independent of $Q$, while
\begin{equation} \label{eq:reflmatrixdR}
\Delta R = \left[ \! \! \begin{array}{cc}
0 & \sin\!{\phi} \sin\!{\theta} r_{sp}^{\parallel} + \cos\!{\theta} r_{sp}^{\perp} \\
-\sin\!{\phi} \sin\!{\theta} r_{sp}^{\parallel} + \cos\!{\theta} r_{sp}^{\perp} & \cos\!{\phi} \sin\!{\theta} \Delta r_{pp}
\end{array}
\! \! \right]
\end{equation}
depends on the magnetization. The magnetization orientation has been written out explicitly in terms of the spherical angles $\theta$ and $\phi$ (the $Z$-axis points into the ferromagnetic plate). The terms $r_{ss}$ and $r_{pp}$ are the Fresnel coefficients for a normal metallic mirror, while the terms $r_{sp}^\perp$, $r_{sp}^\parallel$, and $\Delta r_{pp}$ arise from the polar, longitudinal and transverse Kerr effect respectively and are due to the magnetization. All coefficients can be expressed in terms of the diagonal ($\epsilon_{xx}$) and off-diagonal ($\epsilon_{xy}$) components of the dielectric tensor ~\cite{Metzger1965, Metalidis2002}:
\begin{subequations}
\label{eq:reflcoeffFM}
\begin{eqnarray}
r_{ss}({\rm i}\omega,{\rm i}k_{\perp}) &=&
\frac{k_{\perp}c-\xi}{k_{\perp} c+\xi}
\\
r_{pp}({\rm i}\omega,{\rm i}k_{\perp}) &=&
\frac{\epsilon_{xx}({\rm
i}\omega)k_{\perp}c-\xi}{\epsilon_{xx}({\rm
i}\omega)k_{\perp}c+\xi} \\
r_{sp}^{\perp}({\rm i}\omega,{\rm i}k_{\perp})&=&
\frac{-k_{\perp}
c\,\omega\,\epsilon_{xy}({\rm
i}\omega)}{\left[k_{\perp}c+\xi\right]\left[\epsilon_{xx} ({\rm
i}\omega)k_{\perp}c+\xi\right]},\\
r_{sp}^{\parallel}({\rm i}\omega,{\rm i}k_{\perp}) &=&
\frac{-k_{\perp}c
\sqrt{\omega^{2}-(k_{\perp}c)^{2}} \, \omega\,
\epsilon_{xy}({\rm i}\omega)}
{\left(k_{\perp}c+\xi\right)\left(\epsilon_{xx} ({\rm
i}\omega)k_{\perp}c+\xi\right)\xi}, \\
\Delta r_{pp}({\rm i}\omega,{\rm i}k_{\perp}) &=&
\frac{2\sqrt{\omega^{2}-(k_{\perp}c)^{2}} \, \epsilon_{xy}({\rm
i}\omega) \, (k_{\perp}c)} {\left(\epsilon_{xx}
(i\omega)k_{\perp}c+\xi\right)^{2}},
\end{eqnarray}
\end{subequations}
where we have defined: $\xi=\sqrt{\omega^{2} (\epsilon_{xx}({\rm
i}\omega)-1)+(k_{\perp}c)^{2}}$.

\section{Out-of-Plane Magnetic Anisotropy}
Using the reflection coefficients above, we will now consider the Casimir effect between a non-magnetic isotropic mirror $A$ and a ferromagnetic mirror $B$. For this particular case, mirror $A$ is described by a reflection matrix ${\sf R}_{A} = {\sf R}^0_A$, and mirror $B$ by ${\sf R}_{B} = {\sf R}^0_B + \Delta{\sf R}$ as defined in Eqs.~(\ref{eq:reflmatrixR0}),~(\ref{eq:reflmatrixdR}) and~(\ref{eq:reflcoeffFM}). Making an expansion in $\Delta {\sf R}$ of the logarithm in Eq.~(\ref{eq:CasEnergy}) and keeping only terms in the lowest non-vanishing order, one obtains the following expression for the Casimir energy:
\begin{equation} \label{eq:CasEnergyContributions}
{\cal E} = {\cal E}_0 + {\cal E}^\perp \cos^2\theta + \left({\cal E}^\parallel_1 + {\cal E}^\parallel_2\right) \sin^2\theta.
\end{equation}
In this expression, ${\cal E}_0$ is independent on the magnetization orientation and describes the non-magnetic contribution to the Casimir energy, while
\begin{widetext}
\begin{subequations} \label{eq:CasEnergyContributionsExpressions}
\begin{eqnarray}
{\cal E}^\perp &=& -\frac{\hbar}{4\pi^2} \int_{0}^{+\infty} {\rm d}
k_{\perp}\,k_{\perp} \int_{0}^{k_{\perp}c} {\rm d}\omega\ \frac{r_{ss}^B r_{pp}^B (r_{sp}^\perp)^2 {\rm e}^{-4k_{\perp}D}}{(1-r_{ss}^A r_{ss}^B {\rm e}^{-2k_{\perp}D} ) (1-r_{pp}^A r_{pp}^B {\rm e}^{-2k_{\perp}D})}, \\
{\cal E}^\parallel_1 &=& \frac{\hbar}{8\pi^2} \int_{0}^{+\infty} {\rm d}
k_{\perp}\,k_{\perp} \int_{0}^{k_{\perp}c} {\rm d}\omega\ \frac{r_{ss}^B r_{pp}^B (r_{sp}^\parallel)^2 {\rm e}^{-4k_{\perp}D}}{(1-r_{ss}^A r_{ss}^B {\rm e}^{-2k_{\perp}D} ) (1-r_{pp}^A r_{pp}^B {\rm e}^{-2k_{\perp}D})}, \\
{\cal E}^\parallel_2 &=& -\frac{\hbar}{16\pi^2} \int_{0}^{+\infty} {\rm d}
k_{\perp}\,k_{\perp} \int_{0}^{k_{\perp}c} {\rm d}\omega\ \frac{(r_{pp}^B)^2 (\Delta r_{pp})^2 {\rm e}^{-4k_{\perp}D}}{(1-r_{pp}^A r_{pp}^B {\rm e}^{-2k_{\perp}D})^2}
\end{eqnarray}
\end{subequations}
\end{widetext}
are terms arising respectively from the polar, longitudinal and transverse contributions to the magneto-optical Kerr effect.

The Casimir energy in Eq.~(\ref{eq:CasEnergyContributions}) is of course independent of the in-plane magnetization, because the considered system is invariant with respect to rotations around the normal to the plates. However, the energy depends on the angle $\theta$ between the magnetization and the normal to the plates, i.e., the Casimir effect gives rise to an out-of-plane magnetic anisotropy. It is important to note that the contribution coming from the polar Kerr effect, ${\cal E}^\perp$ is proportional to $\cos^2\theta$ and thus tends to align the magnetization parallel to the plates ($\theta = \pi/2$ or $3\pi/2$), while the longitudinal and transverse contributions ${\cal E}_{1,2}^\parallel$ are proportional to $\sin^2\theta$ and have their energy minimum when the magnetization is perpendicular to the plates ($\theta = 0,\pi$). The physical parameters (dielectric constants) of the plates, and as we will see, the interplate distance, will decide which of these competing terms is dominant.

We have numerically calculated the different contributions for a setup where the non-magnetic plate is made of gold, and the magnetic plate is made of iron. The reflection coefficients in Eq.~(\ref{eq:reflcoeffFM}) will depend on the dielectric tensor of these metals. In order to take the interband transitions properly into account, we have calculated the dielectric tensor from the available optical data, instead of reverting to a simple plasma or Drude model. The diagonal part of the dielectric response function $\epsilon_{xx}(\omega)$ of $\text{Au}$ and $\text{Fe}$, evaluated at real frequencies, is tabulated in Ref.~\cite{Palik1991}. A significant amount of data points is available in the range $10^{-1}\ldots10^4 \ \text{eV}$ for $\text{Au}$ and $10^{-3}\ldots10^4 \ \text{eV}$ for $\text{Fe}$. At lower frequencies, where interband transitions do not play a significant role anymore, we have extrapolated the data with a Drude model using a plasma frequency $\hbar \omega_p = 9 \ \text{eV}$ and a relaxation time $\hbar/\tau = 35 \cdot 10^{-3} \  \text{eV}$ for $\text{Au}$, while $\hbar \omega_p = 3.54 \ \text{eV}$ and $\hbar /\tau = 19 \cdot 10^{-3} \ \text{eV}$ for $\text{Fe}$~\cite{Palik1991}. Optical data for the off-diagonal element of the dielectric tensor $\epsilon_{xy}(\omega)$ of $\text{Fe}$ is rather scarce. We have used data points between $0.1 \ \text{eV}$ and $6 \ \text{eV}$ found in Ref.~\cite{Landolt1992}, and assumed $\epsilon_{xy}(\omega)=0$ out of this range. The dielectric constants $\epsilon_{xx}(\mathrm{i} \omega)$ and $\epsilon_{xy}(\mathrm{i}\omega)$ at imaginary frequencies, which are needed for evaluating the reflection coefficients in Eq.~(\ref{eq:reflcoeffFM}), were then obtained by using the Kramers-Kr\"{o}nig relations. Details of the whole procedure, together with plots of the resulting dielectric tensors can be found in Ref.~\cite{Lambrecht2000} and our earlier work~\cite{Metalidis2002}.

\begin{figure}
\includegraphics[width = 9cm]{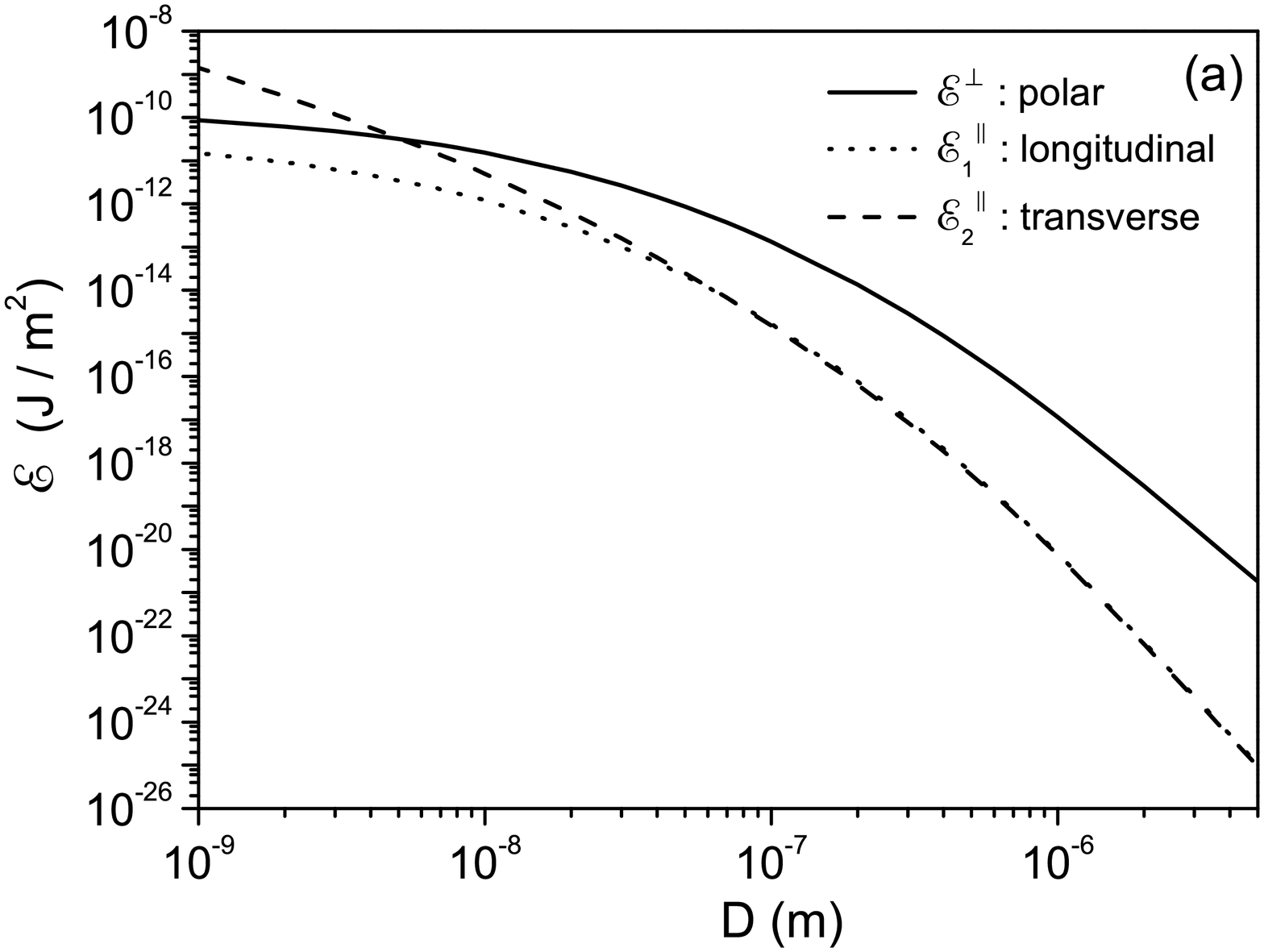}
\vspace*{-0.5cm}
\includegraphics[width = 9cm]{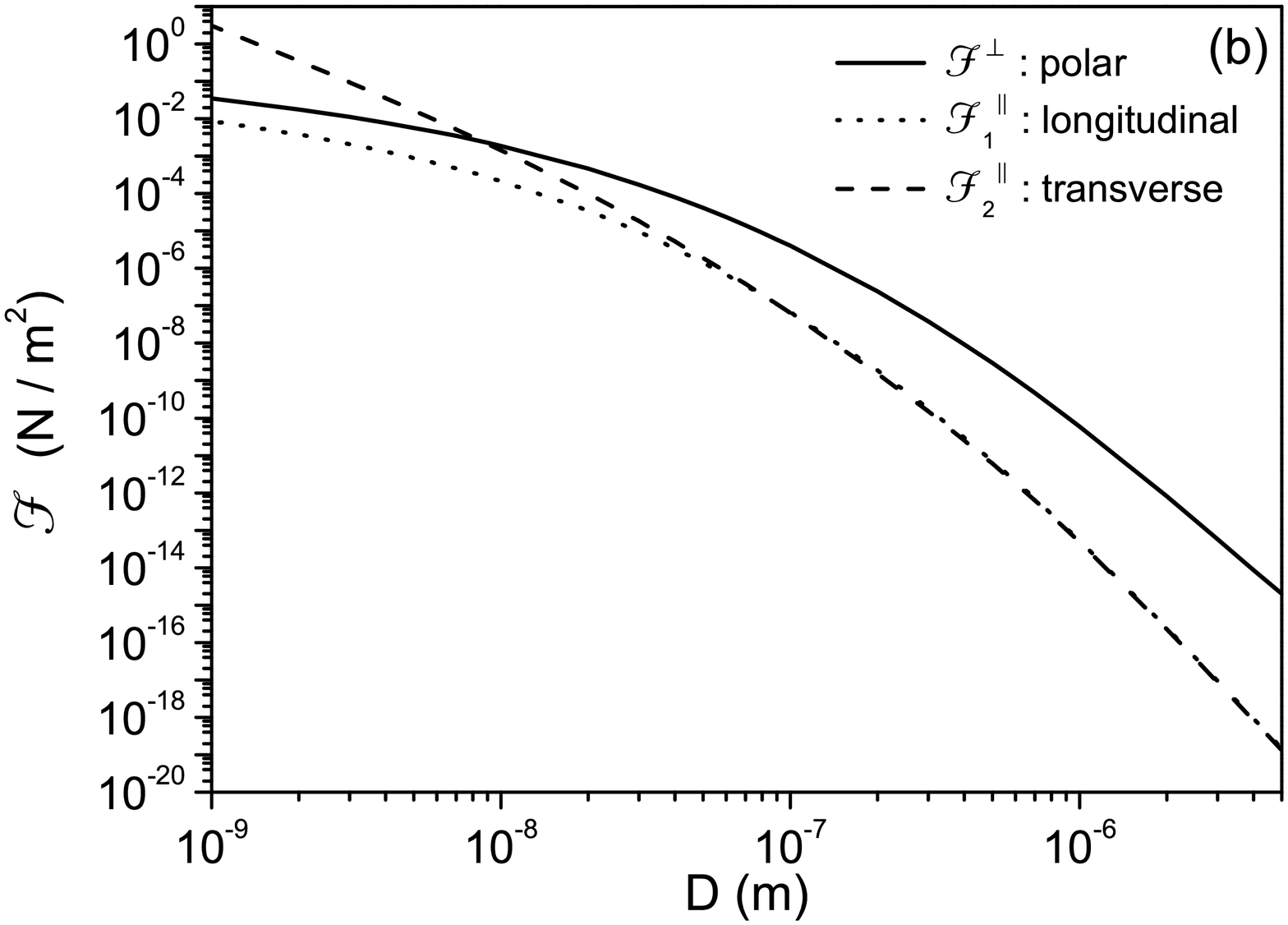}
\caption{\label{fig1} Polar, longitudinal, and transverse contributions to the Casimir energy per unit area (a) and the Casimir force per unit area (b) between an $Fe$ and an $Au$ mirror.}
\end{figure}

Finally, the Casimir energy per unit area is calculated by numerical integration of Eqs.~(\ref{eq:CasEnergyContributionsExpressions}), for interplate distances $D$ ranging from $1 \ \text{nm}$ to $5 \ \mu \text{m}$. Results for the Casimir energy amplitudes ${\cal E}^\perp$, ${\cal E}^\parallel_1$, and ${\cal E}^\parallel_2$ are given in Fig.~\ref{fig1}(a). The corresponding force contributions are obtained by deriving the energy expressions with respect to the interplate distance $D$:
\begin{equation}
{\cal F} = \frac{\partial {\cal E}}{\partial D},
\end{equation}
and are shown in Fig.~\ref{fig1}(b).

For distances larger than approximately $10 \ \text{nm}$, corresponding to the experimentally relevant regime, the contribution ${\cal E}^\perp$ arising from the polar Kerr effect is clearly dominant. Looking back at Eq.~(\ref{eq:CasEnergyContributions}), this means that the lowest energy state at these distances is one where the magnetization is perpendicular to the plates.  However, at smaller distances there is a crossover to a regime where the transverse contribution ${\cal E}^\parallel_2$ dominates, and consequently the preferential magnetization direction will interestingly change from perpendicular towards parallel to the plates upon decreasing the interplate distance. Looking at the distance dependence of the force (per unit area) in Fig.~\ref{fig1}(b), we have found that for long distances the polar contribution varies like $D^{-13/2}$, compared to $D^{-8}$ for the other contributions. For very short distances, the transverse force varies like $D^{-3}$, while the polar and longitudinal terms show (approximately) a $D^{-1}$ dependence.

In order to estimate the Casimir force in a typical experiment, we consider two disks of radius $R = 10 \ \mu\text{m}$, at a distance $D = 100 \ \text{nm}$ from each other. For this setup, we obtain a force amplitude $F = {\cal F} \pi R^2$ on the order of a few fN. The anisotropy could thus possibly be detected with state-of-the-art mechanical (force) measurement schemes ($\text{aN}$ resolution is reached in magnetic resonance force microscopy~\cite{Stowe1997, Rugar1994, Rugar2004}), but presumably not via some magnetic measurement because the anisotropy contribution due to the Casimir effect is considerably smaller (by several orders of magnitude) than the shape anisotropy and the magnetocrystalline surface anisotropy of the $\text{Fe}$ plate. We have therefore looked at a different configuration where the latter anisotropies are essentially zero. In particular, we considered a setup where the magnetic mirror is now placed opposite of a non-magnetic mirror with uniaxial (optical) anisotropy. The latter can, e.g., be made from a uniaxial birefringent material with the optical axis in the plane. For this kind of system, we have found that the Casimir effect will create an in-plane magnetic anisotropy, which could be much easier to detect. Our results will be presented in the next section.

\section{In-Plane Anisotropy}
The reflection coefficients of a mirror with uniaxial optical anisotropy can be expressed in terms of
the ordinary ($\epsilon_o$) and extraordinary ($\epsilon_e$) dielectric constants of the anisotropic material. The corresponding equations can be found, e.g., in Refs.~\cite{Sosnowski1972, Lekner1991}. We will consider a mirror with the optical axis in the plane, for which one obtains (at imaginary frequency and wavevector)~\cite{Lekner1991}:
\begin{subequations}
\begin{eqnarray}
D' r_{ss}^\text{uni} &=& \left(k_\perp c - \xi_o\right) \left(\frac{\xi_o}{\epsilon_o k_\perp c} + 1 \right) \xi_\alpha^2 \\ &\ & \ \ \ \ - \left(\xi_e - \xi_o\right) \left[ \xi_\alpha^2 + \frac{\xi_o}{k_\perp c}\left[ \omega^2 \beta^2 - (\alpha k_\perp c)^2 \right]  \right] \nonumber \\
D' r_{pp}^\text{uni} &=& -\left(k_\perp c + \xi_o\right) \left(\frac{\xi_o}{\epsilon_o k_\perp c} - 1 \right) \xi_\alpha^2 \\ & \ & \ \ \ + \left(\xi_e - \xi_o\right) \left[ \xi_\alpha^2 - \frac{\xi_o}{k_\perp c}\left[ \omega^2 \beta^2 - (\alpha k_\perp c)^2 \right]  \right] \nonumber \\
r_{sp}^\text{uni} &=& -r_{ps}^\text{uni} = \frac{2 \alpha \beta \xi_o \left(\xi_o - \xi_e \right) \omega}{D'},
\end{eqnarray}
\end{subequations}
with
\begin{subequations}
\begin{eqnarray}
D' &=& \left(k_\perp c + \xi_o\right) \left(\frac{\xi_o}{\epsilon_o k_\perp c} + 1 \right) \xi_\alpha^2 \nonumber \\ & \ & \ \ \ +\left(\xi_e - \xi_o\right) \left[ \xi_\alpha^2 + \frac{\xi_o}{k_\perp c}\left( \omega^2 \beta^2 + (\alpha k_\perp c)^2 \right)  \right], \nonumber \\
\xi_o &=& \sqrt{\omega^2 \left( \epsilon_o - 1\right) + \left( k_\perp c\right)^2}, \nonumber \\
\xi_e &=& \sqrt{\xi_o^2 + \frac{ \epsilon_e - \epsilon_o}{\epsilon_o} \xi_\alpha^2}, \nonumber \\
\xi_\alpha &=& \sqrt{\left(\epsilon_o - \alpha^2\right) \omega^2 + (\alpha k_\perp c)^2}, \nonumber
\end{eqnarray}
\end{subequations}
and where $\alpha = \sin \zeta$ and $\beta = -\cos \zeta$, with $\zeta$ the angle between the optical axis and the $X$-axis. The dielectric functions $\epsilon_o$ and $\epsilon_e$ are evaluated at imaginary frequencies. Please note that in comparison with the expressions in Ref.~\cite{Lekner1991}, both $r_{pp}$ and $r_{ps}$ have a minus sign. This results from a different convention in Ref.~\cite{Lekner1991}, where the $p$ axis reverses sign upon reflection, whereas we keep this axis unchanged~\cite{Muller1969}. Furthermore, the definition of $\alpha$ and $\beta$ is different because we assume the incidence plane to be the $OYZ$ plane as in Ref.~\cite{Jaekel1991}, whereas this is the $OXZ$ plane in Ref.~\cite{Lekner1991}.
\begin{figure}
\includegraphics[width = 7cm]{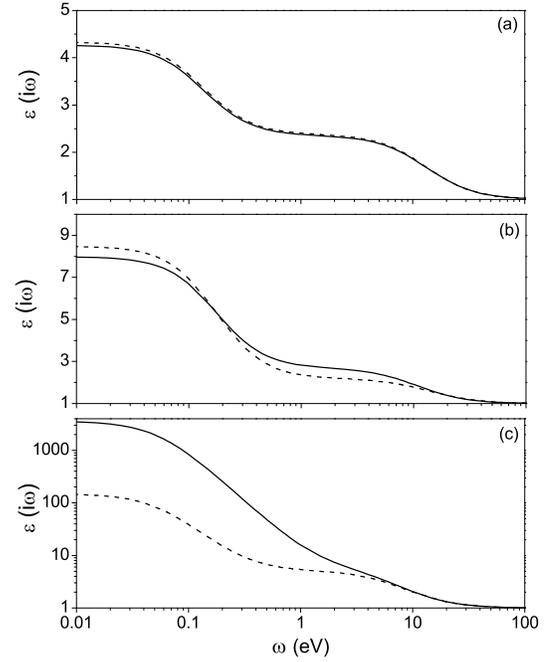}
\caption{\label{fig2} Dielectric functions for quartz~(a), calcite~(b), and barium titanate~(c). Solid and dashed lines correspond to the ordinary and extraordinary dielectric function respectively.}
\end{figure}

It is clear that the angular dependence of these reflection coefficients is much more complicated than what we had for a magnetic mirror in Eq.~(\ref{eq:reflmatrixdR}). Therefore it is more difficult to make analytical progress, and we have opted for a direct numerical evaluation of the Casimir energy in Eq.~(\ref{eq:CasEnergy}). We have considered a setup where one ferromagnetic ($\text{Fe}$) plate with magnetization in the mirror plane is opposed to a non-magnetic uniaxially anisotropic mirror. 

The dielectric tensor of the $\text{Fe}$ plate is calculated directly from optical data, as has been described in the previous section. The uniaxial plate is assumed to be made of one of the birefringent materials quartz ($\text{SiO}_2$), calcite ($\text{CaCO}_3$) or barium titanate ($\text{BaTiO}_3$). Since optical data for these materials is not available in a large enough frequency range, we have used a two-oscillator model to describe their dielectric function. Although this model is quite simple, it gives a rather good description of the dielectric tensor in inorganic materials~\cite{Hough1980, Muller1969}, and has been used already to estimate the Casimir torque between two birefringent plates~\cite{Munday2005}. Within this approach, both the ordinary and extraordinary dielectric function of the uniaxial mirror will be modeled by:
\begin{equation}
\epsilon(\mathrm{i} \omega) = 1 + \frac{C_{IR}}{1+\left( \frac{\omega}{\omega_{IR}}\right)^2} + \frac{C_{UV}}{1+\left( \frac{\omega}{\omega_{UV}} \right)^2},
\end{equation}
where $\omega_{IR(UV)}$ and $C_{IR(UV)}$ are the characteristic absorption frequencies and absorption strengths in the infrared (ultraviolet) range. These constants were calculated in Refs.~\cite{Bergstrom1997, Hough1980} from the available optical data, and were summarized for the ordinary and extraordinary directions of quartz, calcite and barium titanate in Table I of Ref.~\cite{Munday2005}. The resulting dielectric functions are plotted in Fig.~\ref{fig2}~\cite{comment1}. From this figure, one sees that quartz shows the weakest birefringence and has $\epsilon_e > \epsilon_o$ for all frequencies. Barium titanate is the strongest birefringent material and has $\epsilon_e < \epsilon_o$, whereas calcite is somewhere in the middle and shows $\epsilon_e > \epsilon_o$ for small frequencies and $\epsilon_e < \epsilon_o$ for larger frequencies.
\begin{figure}
\includegraphics[width = 9cm]{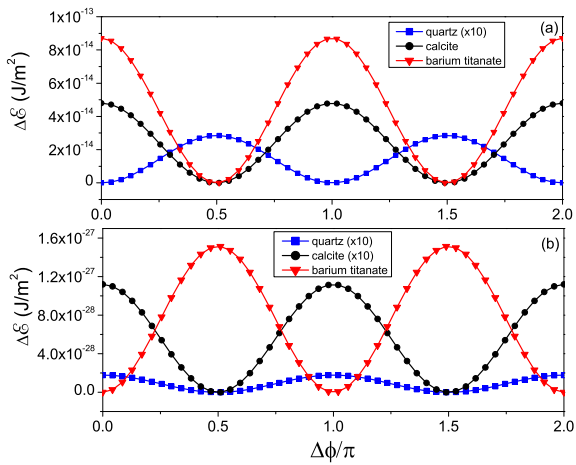}
\caption{\label{fig3} Amplitude of the anisotropy energy (per unit area) for a magnetic Fe mirror placed opposite of a non-magnetic birefringent mirror made out of quartz, calcite or barium titanate. Interplate distance $D=10 \ \text{nm}$~(a) and $D=5 \ \mu\text{m}$~(b). The amplitude is multiplied by 10 for quartz in both figures, and for calcite in~(b).}
\end{figure}

We have calculated the Casimir energy ${\cal E}(\Delta \phi)$ as a function of the angle $\Delta \phi$ between the magnetization of the magnetic mirror (assumed to be in-plane) and the optical axis of the non-magnetic one. Results are shown in Fig.~\ref{fig3} for the three considered birefringent materials, and at distances of $10 \ \text{nm}$ and $5 \ \mu\text{m}$. We have plotted the quantity $\Delta{\cal E} = {\cal E}(\Delta \phi) - {\cal E}_\text{min}$ so that the energy minimum will always lie at zero. All curves could be fitted very well with either a $\sin^2 \Delta \phi$ or a $\cos^2 \Delta \phi$ dependence, as depicted by the solid line connecting the calculated data points.

We observe that at the shortest distance ($10 \ \text{nm}$ in Fig.~\ref{fig3}(a)), the curves for $\text{BaTiO}_3$ and $\text{CaCO}_3$ show a $\cos^2 \Delta \phi$ behavior and thus favor a magnetization that is perpendicular to the optical axis, whereas the curve for $\text{SiO}_2$ varies like $\sin^2 \Delta \phi$ and tends to align the magnetization with the optical axis. This difference is due to the different sign of the quantity $\Delta \epsilon =\epsilon_e - \epsilon_o$ for the three materials in Fig.~\ref{fig2}, and we find that the energy minimum corresponds to a situation where the magnetization is oriented perpendicular to the dielectric axis with the largest value of $\epsilon$. One should note that although $\Delta \epsilon$ changes sign at some intermediate frequency for calcite, the higher frequencies (where $\Delta \epsilon > 0$) will dominate the integrals in Eq.~(\ref{eq:CasEnergy}) at small distances, and calcite therefore shows the same angular dependence as $\text{BaTiO}_3$ in Fig.~\ref{fig3}(a).

Interestingly, at a longer distance ($5 \ \mu\text{m}$ in Fig.~\ref{fig3}(b)), the angular dependence is exactly opposite, at least for quartz and barium titanate: we have a $\sin^2 \Delta \phi$ behavior for $\text{BaTiO}_3$ and a $\cos^2 \Delta \phi$ for quartz (see Fig.~\ref{fig3}) so that now the energy is minimized for a magnetization parallel to the dielectric axis with the largest value of $\epsilon$. In the previous section, we observed a similar change of the angular dependence for a setup with an optically isotropic mirror, and explained it by a competition between the polar and transverse Kerr contributions. Although the polar effect is zero in the setup we consider now (the magnetization lies in the plane, so $\theta = \pi/2$ in Eq.~(\ref{eq:reflmatrixdR})), the two remaining longitudinal and transverse contributions can still compete with each other and will lead to the observed angular crossover. It is however not possible to address both contributions separately, since mixed terms of the form $r_{sp}^\parallel \Delta r_{pp}$ will also appear in the energy expression.

Calcite is a special case, as it has the same $\cos^2 \Delta \phi$ behavior both in the limit of short and long distances (see Fig.~\ref{fig3}). In principle, the competition between longitudinal and transverse Kerr effect contributions would lead to the same crossover as in the other two considered materials. However, at long distances the low frequencies will determine the integrals in Eq.~(\ref{eq:CasEnergy}). As depicted in Fig.~\ref{fig2}(b), one has $\Delta \epsilon < 0$ at these low frequencies, compared to $\Delta \epsilon > 0$ for the higher frequencies which determined the integrals at short distances. It is this sign change of $\Delta \epsilon$ that will cancel the angular dependence crossover due to the Kerr effect competition. Nevertheless, there will be an intermediate distance regime where the higher frequencies with $\Delta \epsilon > 0$ are still the most important, and the competition between the Kerr contributions is decisive in changing the angular dependence from $\cos^2 \Delta \phi$ to $\sin^2 \Delta \phi$.
\begin{figure}
\includegraphics[width = 9cm]{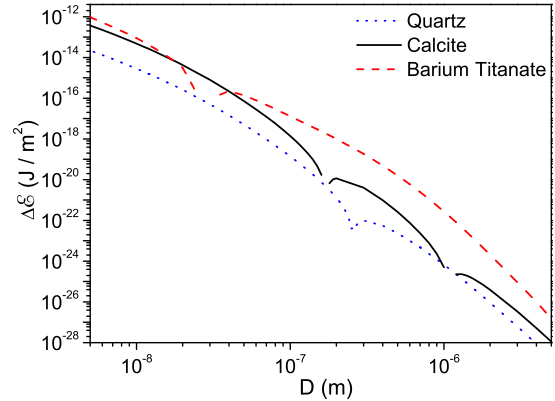}
\caption{\label{fig4} Amplitude of the anisotropy energy (per unit area) as a function of interplate distance $D$ for a magnetic Fe mirror placed opposite of a non-magnetic mirror made out of quartz, calcite, or barium titanate.}
\end{figure}

This becomes clearer in Fig.~\ref{fig4}, where the amplitude of the $\Delta \phi$ oscillations is shown. Both the curves for $\text{BaTiO}_3$ and quartz show a single kink. At the kink, the $\sin^2 \Delta \phi$ and $\cos^2 \Delta \phi$ contributions to the anisotropy energy are of equal magnitude so the net energy amplitude goes to zero. This kink thus signals a crossover point, and as explained above, one clearly observes two such crossovers for calcite but only a single one for quartz and barium titanate. Furthermore we see from Fig.~\ref{fig4} that, as expected, the anisotropy energy is proportional to the amount of birefringence of the considered material, and thus largest for $\text{BaTiO}_3$ and smallest for quartz.

An experimental detection of these anisotropy effects will be quite challenging since the calculated energies are rather small. However, the studied system has the advantage to lend itself to different measurement setups. Firstly, a force measurement could be done by evaporating a droplet of iron on a cantilever and placing it over a birefringent plate. A resonant rf-field can be used to rotate the magnetization, and the Casimir force will result in oscillations at the resonance frequency of the cantilever. For an iron droplet with a curvature radius $R = 100 \ \mu\text{m}$, the force amplitude $F$ can be estimated by using the force proximity theorem: $F = 2 \pi R \Delta {\cal E}$. At a distance of $10 \text{nm}$, the force is $F = 60 \ aN$ for $\text{BaTiO}_3$. Force detection with aN-resolution has already been shown in similar magnetic resonance force microscopy setups~\cite{Rugar1994, Stowe1997, Rugar2004}. Secondly, the anisotropy energy will give rise to a mechanical torque $\partial \Delta {\cal E} / \partial \Delta \phi$ that seeks to rotate the plates towards their minimal energy position. For plates with a radius of $R=100 \ \mu\text{m}$, the torque at an interplate distance $D = 10 \ \text{nm}$ is $\approx 10^{-21} \ \text{Nm}$. A possible setup to demonstrate such torques can be found in Ref.~\cite{Munday2005}, where both plates are immersed in a liquid one on top of the other. With a suitable selection of material parameters, a repulsive Casimir force will balance gravity and a birefringent disk can be held floating above a hard ferromagnetic plate. The Casimir torque will then rotate the plates back to their minimum energy orientation once they are brought out of their equilibrium position. A third measurement scheme would consist of detecting the magnetic anisotropy energy of the ferromagnetic plate directly with a magnetic measurement. By using a cylindrically shaped ferromagnetic mirror cut along the correct axis, both the shape and magnetocrystalline anisotropy can be made small. Furthermore, no anisotropy contribution intrinsic to the ferromagnetic plate will depend on the orientation of the birefringent plate, so that parisitic effects can in general be filtered out efficiently. Finally, we would like to mention that instead of using a birefringent material such as barium titanate to create the uniaxially anisotropic mirror, the optical anisotropy can also be realized by means of some appropriate nanostructuring, e.g., by creating parallel stripes on an isotropic gold mirror. The optical anisotropy can then be made much larger, thereby increasing the magnetic anisotropy effects (a factor of $1000$ is possible~\cite{Rodrigues2006}). A detailed calculation for this case is not at all straightforward, and beyond the scope of the present paper.

\section{Conclusion}
We have considered the Casimir effect between a ferromagnetic and a non-magnetic mirror. If the non-magnetic mirror is optically isotropic, the Casimir effect is shown to give rise to an out-of-plane magnetic anisotropy in the ferromagnetic plate. For a non-magnetic mirror with uniaxial anisotropy, we found an in-plane anisotropy. In both cases, the angular position with the lowest energy depends on the distance between the plates and is determined by the competition between different contributions to the magneto-optical Kerr effect. Different setups for measuring the effect have been proposed.

\end{document}